# Fifty Years of Transaction Processing Research (extended)[1]

Philip A. Bernstein
Microsoft Research
Redmond, WA, U.S.A.
philbe@microsoft.com

## Abstract

In this short paper, I recount some early history of transaction research (including some of my own), explain why transaction research continues to this day (even though it seems to be a solved problem), and speculate about its future.

**Early history.** Fifty years ago, Jim Gray and his IBM colleagues published their first of a sequence of papers that defined the transaction abstraction and mechanisms to support it: two-phase locking (2PL) for isolation [41][50] and logging for atomicity and durability. I recently summarized many of Gray's other transaction papers in [6]. For over a decade, Gray's 1978 "Notes on Database Operating Systems" was the best explanation of how to implement transactions [48]. Later, when Gray and Andreas Reuter published their book [51], the transaction problem seemed to be solved. Yet research continues, for good reasons.

One driver of systems-oriented database research is algorithmic optimization of existing mechanisms. The ARIES algorithm for logging is a good example [79]. It made a compelling argument for performing a redo pass before an undo pass and explained compensation log records (CLRs) and many other subtleties. Another is the TPC-A/B benchmark, which led to many algorithmic optimizations, since a winning solution often had to replace an inefficient algorithm (e.g., that flushes all dirty pages before commit) with a faster one (e.g., that only flushes the log).

From the beginning, it was recognized that relaxing correctness could lead to better performance. Weaker isolation levels, notably read-committed, offer over 3x higher throughput than 2PL [25]. Why users accept it is a mystery. Perhaps the anomalous behavior it allows does not happen often, or no one notices the errors, or databases have so many errors for other reasons that this source of errors can be ignored. Other widely supported weaker isolation levels are cursor stability and repeatable reads [32], and especially snapshot isolation [5].

**New platforms and mechanisms** are a second driver of systems-oriented database research. In the late 1970's researchers turned their attention to distributed on-line transaction processing (OLTP) (a new platform), typically in a shared-nothing architecture. Since each server has its own log, a distributed update transaction requires the two-phase commit (2PC) protocol, published a few years earlier by Lampson and Sturgis [67][68]. Tandem's NonStop SQL was an early commercial shared-nothing OLTP database management system (DBMS) [95](coauthored by Jim Gray).

Unfortunately, 2PC blocks the termination of a transaction at a participant, P, if a failure prevents P from learning the commit/abort decision. This leads to the CAP conjecture, which says you can get only two of consistency, availability, or partition tolerance [44][87]. It was later proved in [47]. Nevertheless, some consistency and isolation levels can be supported with high availability, as shown in [3].

The alternative to shared-nothing is data sharing, where multiple servers can update the same database. It uses a global lock manager to coordinate cache coherence between servers (by ensuring at most one server has write-access to a page) and ensures log entries by different servers for the same page are correctly ordered. Early implementations were Digital Equipment Corp.'s (now Oracle's) Rdb/VMS on VAXcluster and IBM DB2 Data Sharing. Concurrency control and recovery require rather different algorithms than shared nothing, as shown by Lomet [76] and Mohan and Narang [81][82].

In the mid-1980's, disk prices declined sufficiently to make data replication economically feasible, first as mirrored disks, and later as network latency and bandwidth improved, as remote replicas. Cheaper disks also reduced the cost of multi-version concurrency control [86], which avoids interference between long-running queries and update transactions.

Replication algorithms first appeared in the 1970's. The primary copy approach allows updates only on the primary copy of each replicated data item and pipelines them in serialization order to replicas [93]. Multi-master replication allows a transaction to update any copy, and conflicting updates that executed independently on different replicas are merged [60]. In a highly influential paper, Robert Thomas solved the problem using two innovations, majority consensus and transaction timestamps [101]. Later solutions used vector clocks [42][43][64][83][100].

Optimistic concurrency control (OCC) was invented by Kung and Robinson [63] and independently by Gawlick and Kinkade [46]. Though 2PL outperforms OCC for high-conflict workloads, OCC has lower overhead in low-conflict workloads, and can be easier to implement, especially in distributed systems.

**My early work.** Starting in 1977, as a consultant to Computer Corporation of America, I worked on the SDD-1 distributed DBMS prototype, which used timestamp-based concurrency control [23]. It assigned a timestamp to each transaction and ensured that conflicting accesses to each data item executed in timestamp order. We thought it would out-perform locking because it would require fewer messages, but that turned out to be wrong. A more impactful contribution of SDD-1 was semijoin-based query optimization, which is widely used today. It was proposed by Eugene Wong and fleshed out by Nathan Goodman, Dah-Ming Chiu, and me [10] [17].

Concurrency control algorithms were a hot topic in the 1970's-1980's. Goodman and I saw a lot of similarity between published algorithms. In 1980 we showed that each algorithm was a combination of two techniques drawn from a small repertoire, one for read-write synchronization and one for write-write synchronization [13]. We did a more complete analysis of this in our 1981 survey [14]. We defined one-copy serializability as a correctness standard for multi-copy data [2], and used it to analyze multiversion concurrency control (MVCC) in [15] and replication in [16]. We expanded this line of work in our book [19]. Vassos Hadzilacos joined our book effort, greatly improving the theoretical analyses of concurrency control and recovery. He also wrote the chapter on 2PC and 3PC. Prof. Y.C. Tay wrote the performance section of the 2PL chapter which summarized some of the results in his Ph.D. thesis [96][97].

While working on OLTP products at Sequoia Systems and Digital Equipment Corporation (both are gone but not forgotten), I learned why a transactional DBMS requires a distributed computing front end (a.k.a. OLTP Monitor). It supports terminal management, multi-threading, and remote procedure call, which were missing from operating systems in those days. (This was well understood by Jim Gray and many others in 1979. I am a slow learner.) This led me to study non-transactional synchronization problems solved by persistent queues [20] and to join an OLTP Monitor standardization effort [18]. I summarized this work in survey papers about OLTP monitors in [8] and middleware in [9], and in a book on transaction processing with Eric Newcomer [21] targeted for practitioners.

**The Cloud.** In the early 1990's, the research community mostly regarded transactions as a solved problem—2PL, 2PC, and ARIES—so research quieted down. That changed in the mid 2000's with the arrival of cloud computing as a new platform. Its distinguishing features are disaggregated storage and elastic compute, which enable dynamically adding storage and compute resources on demand, known as scaling out. Web-scale workloads initially caused people to dismiss transactions as unscalable and unnecessary in the cloud, so some systems were designed to offer only weaker consistency features, such as eventual consistency [33][106]. However, they made a comeback due to developer demand. This led to a variety of system architectures for a scalable transaction system in the cloud.

Early cloud database systems adopted a shared-nothing design, where storage is sharded (i.e., partitioned) and each transaction can update data in only one shard. This restricted programming model allows the system to scale out while avoiding the need for 2PC [11][29]. Each shard is usually replicated using the primary-copy approach, for fault tolerance, high availability, and scaling out reads.

Another replication approach is deterministic transactions, where each request to run a transaction is sent to all database replicas [102][103]. The programming model is constrained to ensure all executions of a transaction produce the same result, and synchronization ensures that the transactions execute in the same order on each replica.

Later shared-nothing systems discarded the restriction that a transaction updates only one shard, notably Google's Spanner [31]. It uses 2PC over the primary copy of shards, each of which is replicated using Paxos [66][84], thereby mitigating 2PC's blocking problem due to failures.

A distinguishing feature of Spanner is its support for external consistency, by which they mean that if a transaction $T_2$ starts after $T_1$ finishes committing, then $T_2$'s timestamp is greater than $T_1$'s timestamp [26]. External consistency is closely related to linearizability [57]. An execution is linearizable if there is an equivalent serial execution of the same transactions such that for every two transactions $T_1$ and $T_2$, if $T_1$ committed before $T_2$ started, then $T_1$ precedes $T_2$ in the serial execution. We proved 2PL ensures linearizability in [24], though we did not use the term "linearizability", which was coined many years later. Note that linearizability does not refer to timestamps. An execution can be linearizable but not externally consistent, and vice versa.

Spanner ensures external consistency using its TrueTime API and accurate hardware clocks. Tamer Eldeeb collaborated with me and others to develop a mechanism using a software-based clock to ensure external consistency [38][40]. The TAPIR system stores read timestamps and rejects a write that would invalidate a previously executed read. It also parallelizes writes to all replicas instead of using a primary copy [110]. Kulkarni et al. [62] proposed hybrid logical clocks (HLC), encoding a logical counter and physical-clock value into a single 64-bit timestamp that preserves Lamport's happened-before relation [65] while staying within bounded drift of wall-clock time.

FoundationDB uses a different design using OCC. It centralizes a timestamp server, uses MVCC, shards the OCC

validators and storage servers, and avoids 2PC by having each transaction write to a single replicated log [112].

Although the first cloud database systems were shared nothing, data sharing systems soon followed [1][104][105]. Instead of servers sharing disk drives like the 1990's on-premises versions, they shared data in cloud storage. This design is often called *cloud native*. Currently, most such systems allow only one server to perform updates. However, cloud DBMS products from Alibaba and Huawei do allow multiple compute servers to perform updates [73][108]. Other designs for cloud DBMSs appear in [54][70][71][72][78][113].

**Other Topics.** There have been proposals for integrating transactions into the application programming language, such as Argus [74] and Avalon [34][56]. OLTP monitors have offered a shallower integration that consists of a client wrapper over system mechanisms [8][18][92].

Storage latency is an enemy of transaction performance, because it increases the probability of conflict (e.g., due to lock holding time). Main memory DBMSs are a longstanding antidote that is periodically revisited [35][36][59], but they are usually not cost effective [77]. For distributed transactions, network latency is another enemy, because it delays detecting and arbitrating conflicting operations. RDMA helps by reducing average and tail message latency and hence is useful for OLTP DBMSs [27][73][108]. Main memory and RDMA have been combined in some new designs [37][107][109][111][114].

Transaction performance can be improved by splitting a transaction into a multi-step workflow [45][58][84][89][94]. This loses the atomicity and isolation benefits of transactions, which must then be enforced by the application [53].

Many researchers have explored consistency levels other than serializability, such as causal consistency [65][75], eventual consistency [4][105], eventually consistent transactions [28], timeline consistency [30], session consistency [99], parallel snapshot isolation [91], and prefix consistency [98], which are compared in [12].

All these consistency models focus on valid event orderings. An alternative model of consistency is defined [55] in terms of program outcomes. It shows that correctness properties that are monotonic can be preserved with no coordination at all.

**My More Recent Work.** In 2010 I started working with Colin Reid on his proposal for a novel data sharing system, called Hyder, that leverages flash storage (before SSDs were common). In Hyder, the log is the database; each transaction executor replays the log and uses OCC to determine which transactions commit and abort [22]. Since there is only one log, 2PC is not needed.

In 2015, I worked with Tamer Eldeeb on adding transactions to Microsoft's distributed actor system, called Orleans [39]. In Orleans, storage is a plug-in service and often has high latency. We used early-lock release, whereby a transaction T releases its write locks before its commit record is durable [35]. This allows later transactions to read T's dirty output, essentially pipelining their execution to increase throughput.

I worked with Zhihan Guo and Xiangyao Yu on a 2PC variation that reduces storage writes and avoids blocking [52]. In a cloud-based shared-nothing system, all transaction executors can access all storage servers. Therefore, the 2PC coordinator can avoid logging the decision because each transaction executor can tell if all storage servers persisted a transaction's results. Also, by ensuring each participant's log accepts only the first append—either a Prepare or an Abort—the protocol can avoid blocking. There have been many other proposals to improve 2PC to reduce message latency [61][69][88], reduce storage writes [80], or avoid blocking if a coordinator crashes before notifying participants of the final decision [49][90].

**Conclusion** An ideal transaction system executes all transactions with low overhead, maximizing throughput and minimizing latency, and with high availability. It satisfies the following goals:

- Supports single-shard and multi-shard transactions in a distributed database system
- Ensures short and long update transactions perform well
- Ensures low-conflict rate and high-conflict rate workloads perform well
- Supports long-running linearizable queries on consistent snapshots without disadvantaging update transactions
- Scales out with linear performance improvement as resources are added
- Supports geo-distribution of all the above workloads, allowing multi-master updates for high availability
- Supports multiple isolation levels: serializable, snapshot isolated, or weaker (trading correctness for performance)
- Supports stored procedures and interactive client-side transactions
- Avoids constraints on the programming model

Regrettably, the above goals are often in conflict, leading to tradeoffs. Combined with continual platform changes and different optimization goals, they offer a never-ending opportunity for transaction research.

This paper focused on the early work on transaction processing, my own work on the topic, and selected papers that are closely related to my work. The references are intended to be entry points to learn more about these topics, not as a comprehensive survey. My apologies to authors of the many important contributions that I omitted.

## References

[1] Panagiotis Antonopoulos, Alex Budovski, Cristian Diaconu, Alejandro Hernandez Saenz, Jack Hu, Hanuma Kodavalla, Donald Kossmann, Sandeep Lingam, Umar Farooq Minhas, Naveen Prakash, Vijendra Purohit, Hugh Qu, Chaitanya Sreenivas Ravella, Krystyna Reisteter, Sheetal Shrotri, Dixin Tang, Vikram Wakade: Socrates: The New SQL Server in the Cloud. SIGMOD Conference 2019: 1743-1756

[2] Rony Attar, Philip A. Bernstein, Nathan Goodman: Site Initialization, Recovery, and Backup in a Distributed Database System. IEEE Trans. Software Eng. 10(6): 645-650 (1984)

[3] Peter Bailis, Aaron Davidson, Alan D. Fekete, Ali Ghodsi, Joseph M. Hellerstein, Ion Stoica: Highly Available Transactions: Virtues and Limitations. Proc. VLDB Endow. 7(3): 181-192 (2013)

[4] Peter Bailis, Ali Ghodsi: Eventual consistency today: limitations, extensions, and beyond. Commun. ACM 56(5): 55-63 (2013)

[5] Hal Berenson, Philip A. Bernstein, Jim Gray, Jim Melton, Elizabeth J. O'Neil, Patrick E. O'Neil: A Critique of ANSI SQL Isolation Levels. SIGMOD Conference 1995: 1-10

[6] Philip A. Bernstein. 2023. Eight Transaction papers by Jim Gray, https://arxiv.org/abs/2310.04601.

[7] Philip A. Bernstein, Fifty Years of Transaction Processing Research, SIGMOD Conference 2025: 1-2.

[8] Philip A. Bernstein: Transaction Processing Monitors. Commun. ACM 33(11): 75-86 (1990)

[9] Philip A. Bernstein: Middleware: A Model for Distributed System Services. Commun. ACM 39(2): 86-98 (1996)

[10] Philip A. Bernstein, Dah-Ming W. Chiu: Using Semi-Joins to Solve Relational Queries. J. ACM 28(1): 25-40 (1981)

[11] Philip A. Bernstein, Istvan Cseri, Nishant Dani, Nigel Ellis, Ajay Kalhan, Gopal Kakivaya, David B. Lomet, Ramesh Manne, Lev Novik, Tomas Talius: Adapting Microsoft SQL Server for Cloud Computing. ICDE 2011: 1255-1263

[12] Philip A. Bernstein, Sudipto Das: Rethinking eventual consistency. SIGMOD Conference 2013: 923-928

[13] Philip A. Bernstein, Nathan Goodman: Timestamp-Based Algorithms for Concurrency Control in Dist'd Database Systems. VLDB 1980: 285-300

[14] Philip A. Bernstein, Nathan Goodman: Concurrency Control in Distributed Database Systems. ACM Comput. Surv. 13(2): 185-221 (1981)

[15] Philip A. Bernstein, Nathan Goodman: Multiversion Concurrency Control - Theory and Algorithms. ACM Trans. Database Syst. 8(4): 465-483 (1983)

[16] Philip A. Bernstein, Nathan Goodman: An Algorithm for Concurrency Control and Recovery in Replicated Distributed Databases. ACM Trans. Database Syst. 9(4): 596-615 (1984)

[17] Philip A. Bernstein, Nathan Goodman, Eugene Wong, Christopher L. Reeve, James B. Rothnie Jr.: Query Processing in a System for Distributed Databases (SDD-1). ACM Trans. Database Syst. 6(4): 602-625 (1981)

[18] Philip A. Bernstein, Per O. Gyllstrom, Tom Wimberg: STDL - A Portable Language for Transaction Processing. VLDB 1993: 218-229

[19] Philip A. Bernstein, Vassos Hadzilacos, Nathan Goodman: *Concurrency Control and Recovery in Distributed Database Systems*, Addison-Wesley, 1987

[20] Philip A. Bernstein, Meichun Hsu, Bruce Mann: Implementing Recoverable Requests Using Queues. SIGMOD Conference 1990: 112-122

[21] Philip A. Bernstein, Eric Newcomer: Principles of Transaction Processing for Systems Professionals. Morgan Kaufmann (2nd edition), 2009, ISBN 978-1558606234

[22] Philip A. Bernstein, Colin W. Reid, Sudipto Das: Hyder - A Transactional Record Manager for Shared Flash. CIDR 2011: 9-20

[23] Philip A. Bernstein, David W. Shipman, James B. Rothnie Jr.: Concurrency Control in a System for Distributed Databases (SDD-1). ACM Trans. Database Syst. 5(1): 18-51 (1980)

[24] Philip A. Bernstein, David W. Shipman, Wing Shing Wong: Formal Aspects of Serializability in Database Concurrency Control. IEEE Trans. Software Eng. 5(3): 203-216 (1979)

[25] Paul M. Bober, Michael J. Carey: On Mixing Queries and Transactions via Multiversion Locking. ICDE 1992: 535-545

[26] Eric Brewer: Spanner, TrueTime and the CAP Theorem. https://research.google/pubs/spanner-truetime-and-the-cap-theorem. 2017.

[27] Matthew Burke, Sowmya Dharanipragada, Shannon Joyner, Adriana Szekeres, Jacob Nelson, Irene Zhang, Dan R. K. Ports: PRISM: Rethinking the RDMA Interface for Distributed Systems. SOSP 2021: 228-242

[28] Sebastian Burckhardt, Daan Leijen, Manuel Fähndrich, Mooly Sagiv: Eventually Consistent Transactions. ESOP 2012: 67-86

[29] Fay Chang, Jeffrey Dean, Sanjay Ghemawat, Wilson C. Hsieh, Deborah A. Wallach, Michael Burrows, Tushar Chandra, Andrew Fikes, Robert Gruber: Bigtable: A Distributed Storage System for Structured Data. OSDI 2006: 205-218

[30] Brian F. Cooper, Raghu Ramakrishnan, Utkarsh Srivastava, Adam Silberstein, Philip Bohannon, Hans-Arno Jacobsen, Nick Puz, Daniel Weaver, Ramana Yerneni: PNUTS: Yahoo!'s hosted data serving platform. Proc. VLDB Endow. 1(2): 1277-1288 (2008)

[31] James C. Corbett, Jeffrey Dean, Michael Epstein, Andrew Fikes, Christopher Frost, J. J. Furman, Sanjay Ghemawat, Andrey Gubarev, Christopher Heiser, Peter Hochschild, Wilson C. Hsieh, Sebastian Kanthak, Eugene Kogan, Hongyi Li, Alexander Lloyd, Sergey Melnik, David Mwaura, David Nagle, Sean Quinlan, Rajesh Rao, Lindsay Rolig, Yasushi Saito, Michal Szymaniak, Christopher Taylor, Ruth Wang, Dale Woodford: Spanner: Google's Globally Distributed Database. ACM Trans. Comput. Syst. 31(3): 8 (2013)

[32] C. J. Date: Locking and Recovery in a Shared Database System: An Application Programming Tutorial. VLDB 1979: 1-15

[33] Giuseppe DeCandia, Deniz Hastorun, Madan Jampani, Gunavardhan Kakulapati, Avinash Lakshman, Alex Pilchin, Swaminathan Sivasubramanian, Peter Vosshall, Werner Vogels: Dynamo: Amazon's Highly Available Key-value Store. SOSP 2007: 205-220

[34] David Detlefs, Maurice Herlihy, Jeannette M. Wing: Inheritance of Synchronization and Recovery Properties in Avalon/C++. Computer 21(12): 57-69 (1988)

[35] David J. DeWitt, Randy H. Katz, Frank Olken, Leonard D. Shapiro, Michael Stonebraker, David A. Wood: Implementation Techniques for Main Memory Database Systems. SIGMOD Conference 1984: 1-8

[36] Cristian Diaconu, Craig Freedman, Erik Ismert, Per-Åke Larson, Pravin Mittal, Ryan Stonecipher, Nitin Verma, Mike Zwilling: Hekaton: SQL server's memory-optimized OLTP engine. SIGMOD Conference 2013: 1243-1254

[37] Aleksandar Dragojevic, Dushyanth Narayanan, Miguel Castro, Orion Hodson: FaRM: Fast Remote Memory. NSDI 2014: 401-414

[38] Tamer Eldeeb, Philip A. Bernstein, Asaf Cidon, Junfeng Yang: Chablis: Fast and General Transactions in Geo-Distributed Systems. CIDR 2024, https://www.cidrdb.org/cidr2024/

[39] Tamer Eldeeb, Sebastian Burckhardt, Reuben Bond, Asaf Cidon, Junfeng Yang, Philip A. Bernstein: Cloud Actor-Oriented Database Transactions in Orleans. Proc. VLDB Endow. 17(12): 3720-3730 (2024)

[40] Tamer Eldeeb, Xincheng Xie, Philip A. Bernstein, Asaf Cidon, Junfeng Yang: Chardonnay: Fast and General Datacenter Transactions for On-Disk Databases. OSDI 2023: 343-360

[41] Kapali P. Eswaran, Jim Gray, Raymond A. Lorie, Irving L. Traiger: The Notions of Consistency and Predicate Locks in a Database System. Commun. ACM 19(11): 624-633 (1976)

[42] Colin J. Fidge.: Timestamps in message-passing systems that preserve the partial ordering. In K. Raymond (ed.). Proceedings of the 11th Australian Computer Science Conference (ACSC'88). Vol. 10. pp. 56–66.

[43] Michael J. Fischer, Alan Michael: Sacrificing Serializability to Attain High Availability of Data. PODS 1982: 70-75

[44] Armando Fox, Eric A. Brewer: Harvest, Yield and Scalable Tolerant Systems. Workshop on Hot Topics in Operating Systems 1999: 174-178

[45] Hector Garcia-Molina, Kenneth Salem: Sagas. SIGMOD Conference 1987: 249-259

[46] Dieter Gawlick, David Kinkade: Varieties of Concurrency Control in IMS/VS Fast Path. IEEE Database Eng. Bull. 8(2): 3-10 (1985)

[47] Seth Gilbert, Nancy A. Lynch: Brewer's conjecture and the feasibility of consistent, available, partition-tolerant web services. SIGACT News 33(2): 51-59 (2002)

[48] Jim Gray: Notes on Data Base Operating Systems. Advanced Course: Operating Systems 1978. Springer, LNCS 60: 393-481

[49] Jim Gray, Leslie Lamport: Consensus on transaction commit. ACM Trans. Database Syst. 31(1): 133-160 (2006)

[50] Jim Gray, Raymond A. Lorie, Gianfranco R. Putzolu, Irving L. Traiger: Granularity of Locks in a Large Shared Data Base. VLDB 1975: 428-451

[51] Jim Gray, Andreas Reuter: Transaction Processing: Concepts and Techniques. Morgan Kaufmann 1993, ISBN 1-55860-190-2

[52] Zhihan Guo, Xinyu Zeng, Kan Wu, Wuh-Chwen Hwang, Ziwei Ren, Xiangyao Yu, Mahesh Balakrishnan, Philip A. Bernstein: Cornus: Atomic Commit for a Cloud DBMS with Storage Disaggregation. Proc. VLDB Endow. 16(2): 379-392 (2022)

[53] Pat Helland, Don Haderle: Engagements: Building Eventually ACiD Business Transactions. CIDR 2013

[54] Pat Helland: Scalable OLTP in the Cloud: What's the BIG DEAL? CIDR 2024

[55] Joseph M. Hellerstein, Peter Alvaro: Keeping CALM: when distributed consistency is easy. Commun. ACM 63(9): 72-81 (2020)

[56] Herlihy, M., and Wing, J. Avalon: Language support for reliable distributed systems. 17th International Symposium on Fault-Tolerant Computing. IEEE, New York, 1987

[57] Maurice Herlihy, Jeannette M. Wing: Linearizability: A Correctness Condition for Concurrent Objects. ACM Trans. Program. Lang. Syst. 12(3): 463-492 (1990)

[58] Sushil Jajodia, Larry Kerschberg (editors): Advanced Transaction Models and Architectures. Kluwer 1997, ISBN 0-7923-9880-7

[59] Robert Kallman, Hideaki Kimura, Jonathan Natkins, Andrew Pavlo, Alex Rasin, Stanley B. Zdonik, Evan P. C. Jones, Samuel Madden, Michael

Stonebraker, Yang Zhang, John Hugg, Daniel J. Abadi: H-store: a high-performance, distributed main memory transaction processing system. Proc. VLDB Endow. 1(2): 1496-1499 (2008)

[60] Leonard Kawell Jr., Steven Beckhardt, Timothy Halvorsen, Raymond Ozzie, Irene Greif: Replicated Document Management in a Group Communication System, Proc. of the 2nd ACM Conference on Computer-Supported Cooperative Work, 1988

[61] Tim Kraska, Gene Pang, Michael J. Franklin, Samuel Madden, Alan D. Fekete: MDCC: multi-data center consistency. EuroSys 2013: 113-126

[62] Sandeep S. Kulkarni, Murat Demirbas, Deepak Madappa, Bharadwaj Avva, Marcelo Leone: Logical Physical Clocks. OPODIS 2014: 17-32

[63] H. T. Kung, John T. Robinson: On Optimistic Methods for Concurrency Control. ACM Trans. Database Syst. 6(2): 213-226 (1981)

[64] Rivka Ladin, Barbara Liskov, Liuba Shrira, Sanjay Ghemawat: Providing High Availability Using Lazy Replication. ACM Trans. Comput. Syst. 10(4): 360-391 (1992)

[65] Leslie Lamport: Time, Clocks, and the Ordering of Events in a Distributed System. Commun. ACM 21(7): 558-565 (1978)

[66] Leslie Lamport: The Part-Time Parliament. ACM Trans. Comput. Syst. 16(2): 133-169 (1998)

[67] Butler W. Lampson: Atomic Transactions. Advanced Course: Distributed Systems 1980, Springer LNCS 105: 246-265, 1980.

[68] Butler W. Lampson, Howard E. Sturgis: Crash recovery in Distributed Data Storage Systems. Tech. Report, Xerox Palo Alto Research Center, 1975.

[69] Butler W. Lampson, David B. Lomet: A New Presumed Commit Optimization for Two Phase Commit. VLDB 1993: 630-640

[70] Justin J. Levandoski, David B. Lomet, Mohamed F. Mokbel, Kevin Zhao: Deuteronomy: Transaction Support for Cloud Data. CIDR 2011: 123-133

[71] Justin J. Levandoski, David B. Lomet, Sudipta Sengupta, Ryan Stutsman, Rui Wang: High Performance Transactions in Deuteronomy. CIDR 2015

[72] Justin J. Levandoski, David B. Lomet, Sudipta Sengupta, Ryan Stutsman, Rui Wang: Multi-Version Range Concurrency Control in Deuteronomy. Proc. VLDB Endow. 8(13): 2146-2157 (2015)

[73] Guoliang Li, Wengang Tian, Jinyu Zhang, Ronen Grosman, Zongchao Liu, Sihao Li: GaussDB: A Cloud-Native Multi-Primary Database with Compute-Memory-Storage Disaggregation. Proc. VLDB Endow. 17(12): 3786-3798 (2024)

[74] Barbara Liskov: Distributed Programming in Argus. Commun. ACM 31(3): 300-312 (1988)

[75] Wyatt Lloyd, Michael J. Freedman, Michael Kaminsky, David G. Andersen: Stronger Semantics for Low-Latency Geo-Replicated Storage. NSDI 2013: 313-328

[76] David B. Lomet: Recovery for Shared Disk Systems Using Multiple Redo Logs. Digital Equipment Corp Technical Report CRL 90/4.

[77] David B. Lomet: Cost/performance in modern data stores: how data caching systems succeed. DaMoN 2018: 9:1-9:10

[78] David B. Lomet: Deuteronomy 2.0: Record Caching and Latch Freedom. CoRR abs/2504.14435 (2025)

[79] C. Mohan, Don Haderle, Bruce G. Lindsay, Hamid Pirahesh, Peter M. Schwarz: ARIES: A Transaction Recovery Method Supporting Fine-Granularity Locking and Partial Rollbacks Using Write-Ahead Logging. ACM Trans. Database Syst. 17(1): 94-162 (1992)

[80] C. Mohan, Bruce G. Lindsay, Ron Obermarck: Transaction Management in the R* Distributed Database Management System. ACM Trans. Database Syst. 11(4): 378-396 (1986)

[81] C. Mohan, Inderpal Narang: Recovery and Coherency-Control Protocols for Fast Intersystem Page Transfer and Fine-Granularity Locking in a Shared Disks Transaction Environment. VLDB 1991: 193-207

[82] C. Mohan, Inderpal Narang: Efficient Locking and Caching of Data in the Multisystem Shared Disks Transaction Environment. EDBT 1992: 453-468

[83] Lev Novik, Irena Hudis, Doug Terry, Sanjay Anand, Vivek J. Jhaveri, Ashish Shah, Yunxin Wu: Peer-to-peer Replication in WinFS, Microsoft Research Technical Report MSR-TR-2006-78.

[84] Brian M. Oki, Barbara Liskov: Viewstamped Replication: A General Primary Copy. PODC 1988: 8-17

[85] Dan Pritchett: BASE: An Acid Alternative. ACM Queue 6(3): 48-55 (2008)

[86] David P. Reed: Implementing Atomic Actions on Decentralized Data. ACM Trans. Comput. Syst. 1(1): 3-23 (1983)

[87] James B. Rothnie Jr., Nathan Goodman: A Survey of Research and Development in Distributed Database Management. VLDB 1977: 48-62

[88] George Samaras, Kathryn Britton, Andrew Citron, C. Mohan: Two-Phase Commit Optimizations and Tradeoffs in the Commercial Environment. ICDE 1993: 520-529

[89] Dennis E. Shasha, Eric Simon, Patrick Valduriez: Simple Rational Guidance for Chopping Up Transactions. SIGMOD Conference 1992: 298-307

[90] Dale Skeen: Nonblocking Commit Protocols. SIGMOD Conference 1981: 133-142

[91] Yair Sovran, Russell Power, Marcos K. Aguilera, Jinyang Li: Transactional storage for geo-replicated systems. SOSP 2011: 385-400

[92] Alfred Z. Spector, Joshua J. Bloch, Dean S. Daniels, Richard Draves, Dan Duchamp, Jeffrey L. Eppinger, Sherri G. Menees, Dean S. Thompson: The Camelot Project. IEEE Database Eng. Bull. 9(3): 23-34 (1986)

[93] Michael Stonebraker: Concurrency Control and Consistency of Multiple Copies of Data in Distributed INGRES. IEEE Trans. Software Eng. 5(3): 188-194 (1979)

[94] Michael Stonebraker, Xinjing Zhou, Peter Kraft, Qian Li: Consistency and Correctness in Data-Oriented Workflow Systems, CIDR 2026

[95] Tandem Database Group: NonStop SQL: A Distributed, High-Performance, High-Availability Implementation of SQL. HPTS 1987: 60-104

[96] Y.C. Tay, Locking Performance in Centralized Databases, Academic Press, 1987.

[97] Y.C. Tay, N. Goodman and R. Suri: Locking performance in centralized databases. ACM Trans. on Database Systems 10, 4(Dec. 1985), 415-462.

[98] Doug Terry: Replicated data consistency explained through baseball. Commun. ACM 56(12): 82-89 (2013)

[99] Douglas B. Terry, Alan J. Demers, Karin Petersen, Mike Spreitzer, Marvin Theimer, Brent B. Welch: Session Guarantees for Weakly Consistent Replicated Data. PDIS 1994: 140-149

[100] Douglas B. Terry: Replicated Data Management in Mobile Computing. In: Synthesis Lectures on Mobile and Pervasive Computing, Morgan & Claypool Publishers, 2008.

[101] Robert H. Thomas: A Majority Consensus Approach to Concurrency Control for Multiple Copy Databases. ACM Trans. Database Syst. 4(2): 180-209 (1979)

[102] Alexander Thomson, Daniel J. Abadi: CalvinFS: Consistent WAN Replication and Scalable Metadata Management for Distributed File Systems. FAST 2015: 1-14

[103] Alexander Thomson, Thaddeus Diamond, Shu-Chun Weng, Kun Ren, Philip Shao, Daniel J. Abadi: Calvin: fast distributed transactions for partitioned database systems. SIGMOD Conference 2012: 1-12

[104] Alexandre Verbitski, Anurag Gupta, Debanjan Saha, Murali Brahmadesam, Kamal Gupta, Raman Mittal, Sailesh Krishnamurthy, Sandor Maurice, Tengiz Kharatishvili, Xiaofeng Bao: Amazon Aurora: Design Considerations for High Throughput Cloud-Native Relational Databases. SIGMOD Conference 2017: 1041-1052

[105] Alexandre Verbitski, Anurag Gupta, Debanjan Saha, James Corey, Kamal Gupta, Murali Brahmadesam, Raman Mittal, Sailesh Krishnamurthy, Sandor Maurice, Tengiz Kharatishvili, Xiaofeng Bao: Amazon Aurora: On Avoiding Distributed Consensus for I/Os, Commits, and Membership Changes. SIGMOD Conference 2018: 789-796

[106] Werner Vogels: Eventually Consistent. ACM Queue 6(6): 14-19 (2008)

[107] Xingda Wei, Jiaxin Shi, Yanzhe Chen, Rong Chen, Haibo Chen: Fast in-memory transaction processing using RDMA and HTM. 87-104

[108] Xinjun Yang, Yingqiang Zhang, Hao Chen, Feifei Li, Bo Wang, Jing Fang, Chuan Sun, Yuhui Wang: PolarDB-MP: A Multi-Primary Cloud-Native Database via Disaggregated Shared Memory. SIGMOD Conference Companion 2024: 295-308

[109] Dong Young Yoon, Mosharaf Chowdhury, Barzan Mozafari: Distributed Lock Management with RDMA: Decentralization without Starvation. SIGMOD Conference 2018: 1571-1586

[110] Irene Zhang, Naveen Kr. Sharma, Adriana Szekeres, Arvind Krishnamurthy, Dan R. K. Ports: Building Consistent Transactions with Inconsistent Replication. ACM Trans. Comput. Syst. 35(4): 12:1-12:37 (2017)

[111] Erfan Zamanian, Carsten Binnig, Tim Kraska, Tim Harris: The End of a Myth: Distributed Transaction Can Scale. Proc. VLDB Endow. 10(6): 685-696 (2017)

[112] Jingyu Zhou, Meng Xu, Alexander Shraer, Bala Namasivayam, Alex Miller, Evan Tschannen, Steve Atherton, Andrew J. Beamon, Rusty Sears, John Leach, Dave Rosenthal, Xin Dong, Will Wilson, Ben Collins, David Scherer, Alec Grieser, Young Liu, Alvin Moore, Bhaskar Muppana, Xiaoge Su, Vishesh Yadav: FoundationDB: A Distributed Unbundled Transactional Key Value Store. SIGMOD Conference 2021: 2653-2666

[113] Tobias Ziegler, Philip A. Bernstein, Viktor Leis, Carsten Binnig: Is Scalable OLTP in the Cloud a Solved Problem? CIDR 2023

[114] Tobias Ziegler, Carsten Binnig, Viktor Leis: ScaleStore: A Fast and Cost-Efficient Storage Engine using DRAM, NVMe, and RDMA. SIGMOD Conference 2022: 685-699